\documentclass[useAMS,usenatbib]{mn2e}
\usepackage{graphicx}
\usepackage{deluxetable}
\usepackage{amssymb}

\def\msun{M$_{\odot}$}

\def\alf{${\alpha}$}
\def\mmin{$M_{\rm{min}}$}
\def\mmax{$M_{\rm{max}}$}

\title[The Mass Distribution of Population III Stars]{The Mass Distribution of Population III Stars}
\author[Fraser et al.]{M. Fraser$^{1,2}$\thanks{E-mails: mf,arc,gil@ast.cam.ac.uk},
A.~R. Casey$^2$,
G. Gilmore$^2$
A. Heger$^{3,4,5}$
C. Chan$^{3}$
\\
$^1$School of Physics, O'Brien Centre for Science North, University College Dublin, Belfield, Dublin 4, Ireland.\\
$^2$Institute of Astronomy, University of Cambridge, Madingley Road, Cambridge CB3 0HA, UK\\
$^3$Monash Centre for Astrophysics, School of Physics and Astronomy, Monash University, 19 Rainforest Walk, Vic 3800, Australia\\
$^4$Department of Physics and Astronomy, Shanghai Jiao-Tong University, CNA, Shanghai 200240, P. R. China\\
$^5$School of Physics and Astronomy, University of Minnesota, Minneapolis, MN 55455, USA
\\}

\begin{document}

\date{Accepted 2017 February 22. Received 2017 February 22; in original form 2015 November 08}

\pagerange{\pageref{firstpage}--\pageref{lastpage}} \pubyear{2015}

\maketitle

\label{firstpage}

\begin{abstract}
Extremely metal-poor stars are uniquely informative on the nature of
massive Population III stars.  Modulo a few elements that vary with
stellar evolution, the present-day photospheric abundances observed in
extremely metal-poor stars are representative of their natal gas cloud
composition.  For this reason, the chemistry of extremely
metal-poor stars closely reflects the nucleosynthetic yields of
supernovae from massive Population III stars.  Here we collate
detailed abundances of $53$ extremely metal-poor stars from
the literature and infer the masses of their Population III
progenitors.  We fit a simple initial mass function to  a subset of 29 of the
inferred Population III star masses, and find that the mass
distribution is well-represented by a power law IMF with exponent
$\alpha = 2.35^{+0.29}_{-0.24}$.  The inferred maximum progenitor
mass for supernovae from massive Population III stars is $M_{\rm{max}}
= 87^{+13}_{-33}$\,\msun, and we find no evidence  in our sample
for a contribution from stars with masses above $\sim$120 \msun.  The minimum mass is
strongly consistent with the theoretical lower mass limit for
Population III supernovae.  We conclude that the IMF for massive
Population III stars is consistent with the initial mass function of
present-day massive stars and there may well have formed stars much
below the supernova mass limit that could have survived to the present
day.
\end{abstract}


\section{Introduction}

Population III stars formed from primordial metal-free gas, and can be
subdivided into Population III.1 and III.2 stars. The former are completely 
unaffected by preceding star-formation, while the latter may have been 
effected by energy input from previous star formation, but not chemically enriched \citep{Osh08}.
Of these stars, only those which had masses below
$\lesssim$0.8 \msun\ could have survived to the present day.  No low
mass Population III star has been discovered thus far, however,
despite decades of intensive searching
\citep[e.g.,][]{Bee05,Christlieb08,Norris13,Jacobson15,Sch2014}.
The absence of an unambiguous detection of a low mass Population III
star means that both the individual and ensemble properties of these
stars remains unknown.  The initial mass function (IMF) which governs
the relative frequency of stars as a function of mass, appears to be
fairly uniform today \citep{Bas10}.  In the early Universe, however,
the conditions for star formation were presumably quite different, and
there have long been theoretical arguments that many more
extremely massive ($\sim10^2-10^3$\msun) stars would have formed
\citep[e.g.,][]{Bro99}.  More recently, this has been questioned, with
increased spatial resolution in simulations of metal-free star
formation and an improved understanding of the role of cooling from
molecular hydrogen.  This work suggests that fragmentation occurs, and
that lower-mass metal-free stars can be produced
\citep[e.g.][]{Cla11,Gre11}.

To determine the IMF for Population III stars, we need to count the
number of massive stars in the early Universe.  Because we cannot
observe individual members of the first generation of massive stars in
the foreseeable future, we are forced to turn to indirect tracers.  One
such probe is spectroscopy of galaxies, which can reveal the signature
of a large number of Population III stars
\citep{Sch02,Sob15}.  Alternatively, one can search for the explosion
of Population III stars as high redshift supernovae
\citep{Wha13}.  Finally, it is also possible to study Population III
stars through their chemical imprint on subsequent generations of low
mass stars \citep{Heg02,Heger10}.

Extremely metal-poor (EMP) stars are thought to have formed out of
material that has been enriched by very few (or just one)
supernovae. 
 While they will burn H to form He, low mass stars will not produce
any heavy elements in their cores, 
and  hence  the abundances measured in EMP
stars from high-resolution spectra generally reflect the composition of the gas
from which that star formed.
If this gas has been polluted by only a
single supernova, then these same abundances provide information on
its progenitor and explosion parameters.  Through comparison to models
of supernova nucleosynthesis, we can infer both the mass of the
progenitor and the supernova explosion energy
\citep[][]{Tom14,Placco15,Ji15}.
One important exception to this is the case of giant
stars, which have evolved off the main sequence. For these stars, dredge-up 
can mix CNO processed material from the core of the star to the surface, 
altering the abundances in these elements. For this reason, we consider the combined 
CNO abundances in these objects (Sect. \ref{sect:fitting}). 

In this Letter we assemble a large sample of EMP stars from the
literature, and use them to determine the mass range of massive
Population III stars and infer the slope of the IMF.  We also consider
whether there is any evidence for pair instability supernovae based on
the chemical abundances of stars in the Milky Way.  In Sections
\ref{sect:data} and \ref{sect:fitting} we outline our literature
sample of observational data, the methodology for determining the mass
of the progenitors of Population III SNe, and the procedure for
fitting the IMF.  In Section \ref{sect:discussion} we discuss our
results and draw conclusions.

\section{Data}
\label{sect:data}

We searched the literature for EMP stars with detailed chemical
abundances.  For stars with progressively higher metallicities, there
is a greater likelihood that the star has been polluted by multiple
generations of previous supernovae.  Indeed, given the right
combination of progenitor mass and explosion energy, a single
primordial supernova can enrich some of the interstellar medium to a
level of ${{\rm [Fe/H]} \sim -3}$ (e.g., \citealt{Chen15}).  For these
reasons we place an arbitrary cut of ${{\rm [Fe/H]} < -3.5}$ on our
sample.  This threshold provides us with a sufficient number of
metal-poor stars with detailed chemical information ($\sim$50), but
limits the number of stars that are more likely to have formed from
multiple supernovae.

We first queried the \texttt{SAGA} database\footnote{Described in
  \citet{Suda08,Suda11} and \citet{Yamada13} and available at
  \texttt{http://saga.sci.hokudai.ac.jp/wiki/doku.php}.  The data were
  retrieved on 2015 August 7.} to create a catalog of stars with metallicities
${{\rm[Fe/H]}<-3.5}$.  We complemented this search with recently
published works from the literature. We sourced measured abundances 
directly from the original literature
(i.e., not from heterogeneous compilations like this work) and
converted the published abundance quantities (typically [X/H] or
[X/Fe]) to $\log_\epsilon({\rm X})$ values using the solar abundance
scale quoted in each paper.  For stars with chemical
abundances published in multiple papers, we favoured more recent
literature sources.
The complete sample of metal-poor stars
used in this work, along with their stellar parameters and the
references to the original literature are listed in
Table~\ref{tab:emp-stars}.

In several cases there were more than one set of chemical abundances
quoted for a single star. In these scenarios we opted for the most
`representative' chemical abundances. Specifically we opted for
chemical abundances measured using $\langle{\rm 3D}\rangle$ model
photospheres, abundances adjusted using trustworthy non-local
thermodynamic equilibrium (LTE) corrections, or abundances measured
from ionised species (over neutral species) wherever possible. These
choices were adopted to minimise systematic effects (e.g., non-LTE
departures), which can be significant in EMP stars.

Our literature search revealed many low-metallicity stars with
abundance measurements for just a few elements.  These stars were
faint (some with $V \sim 20$), and therefore require a significant
8~m-class telescope time investment merely to confirm their metal-poor
nature.  Low S/N and/or limited wavelength coverage further precluded
any detailed chemical analysis of these stars.  This point highlights
the need for apparently bright EMP stars, where the detailed chemistry
can be inferred using existing telescope/instrument combinations
\citep[e.g.,][]{Sch2014,Casey2015}.  Previous studies
\citep[e.g.,][]{Bessell15} have commented on the necessity of C, N,
and other light-element abundances in order to accurately infer
supernova progenitor masses. For these reasons, literature stars with
just a few (typical $\alpha$-capture) elemental abundances were
excluded from this work, as they cannot sufficiently constrain
supernova models.  We further excluded stars where there was ambiguity
in the stellar parameters (and therefore the chemistry), usually where
a star is equally likely to be a dwarf or sub-giant.

\section{Mass and IMF fitting}
\label{sect:fitting}

With the assembled chemical abundances for all metal-poor stars, we
used the {\sc StarFit}\footnote{Available through \texttt{http://www.starfit.org}.} code to infer
progenitor masses, remnant masses and explosion energies.  
StarFit will efficiently find the best-fitting model to an observed chemical abundance pattern by comparison to a large database of nucleosynthesis yields and their associated progenitor properties. The assumption is made that the observational uncertainties are normally distributed in logarithmic space and that the abundance values are uncorrelated. This allows the reduced $\chi^2$ residual to be used as a measure of the quality of fit of a model to observations. The residual is defined as

\begin{equation}
\tilde\chi^2 = \frac{1}{N}
\sum_{i=1}^N
\left\{
\begin{array}{ll} 
\left(\frac{o_i - a_i}{\sigma_i}\right)^2 & \mathrm{if}\ o_i\ \mathrm{is\ a\ detection} \\
-2\ln \Phi \left( - \frac{o_i - a_i}{\sigma_i} \right) & \mathrm{if}\ o_i\ \mathrm{is\ an\ upper\ limit}
\end{array}
\right.
\end{equation}

where $N$ is the number of elements being matched, $o_i$ is the observed abundance of element $i$ in dex, $a_i$ is the model abundance, $\sigma_i$ is the uncertainty in the observation, and $\Phi$ is the cumulative distribution function (CDF) of the normal distribution.
Elements for which the model yields are lower than an observational upper limit do not, in principle, contribute to the residual. There exists, however, an uncertainty on the value of the upper limit, which is also assumed to be normally distributed, and thus the contribution to the residual is integrated over the distribution of the upper limit. For each model in the database, the optimal dilution ratio (i.e. the one that minimises the residual) with pristine Big Bang material is calculated by means of Newton-Raphson iteration. Since we only consider contributions from individual yields, the yield with the lowest residual after dilution is chosen as the best fit.

We used
both the supernova models of \citet{Heger10} and the Pair Instability
models of \citet{Heg02}, and excluded Li abundances as they vary
substantially throughout a star's lifetime.  Similarly, due to CNO
cycling, we opted to fit the sum of C, N, and O abundances
simultaneously wherever possible rather than the individual
abundances.  As recommended by \citet{Heger10}, Sc and Cu were treated
as model lower limits because they have multiple different
nucleosynthetic pathways.  We opted to exclude elemental abundances
past the iron-peak ($Z > 30$) as the nucleosynthetic origin of these
elements in Population III stars is quite uncertain.  Each {\sc StarFit} fit
provides an explosion energy, a progenitor mass, a remnant mass,
strength of mixing during the SN explosion, and a best-fit residual.

\begin{figure*}
   \includegraphics[width=\textwidth]{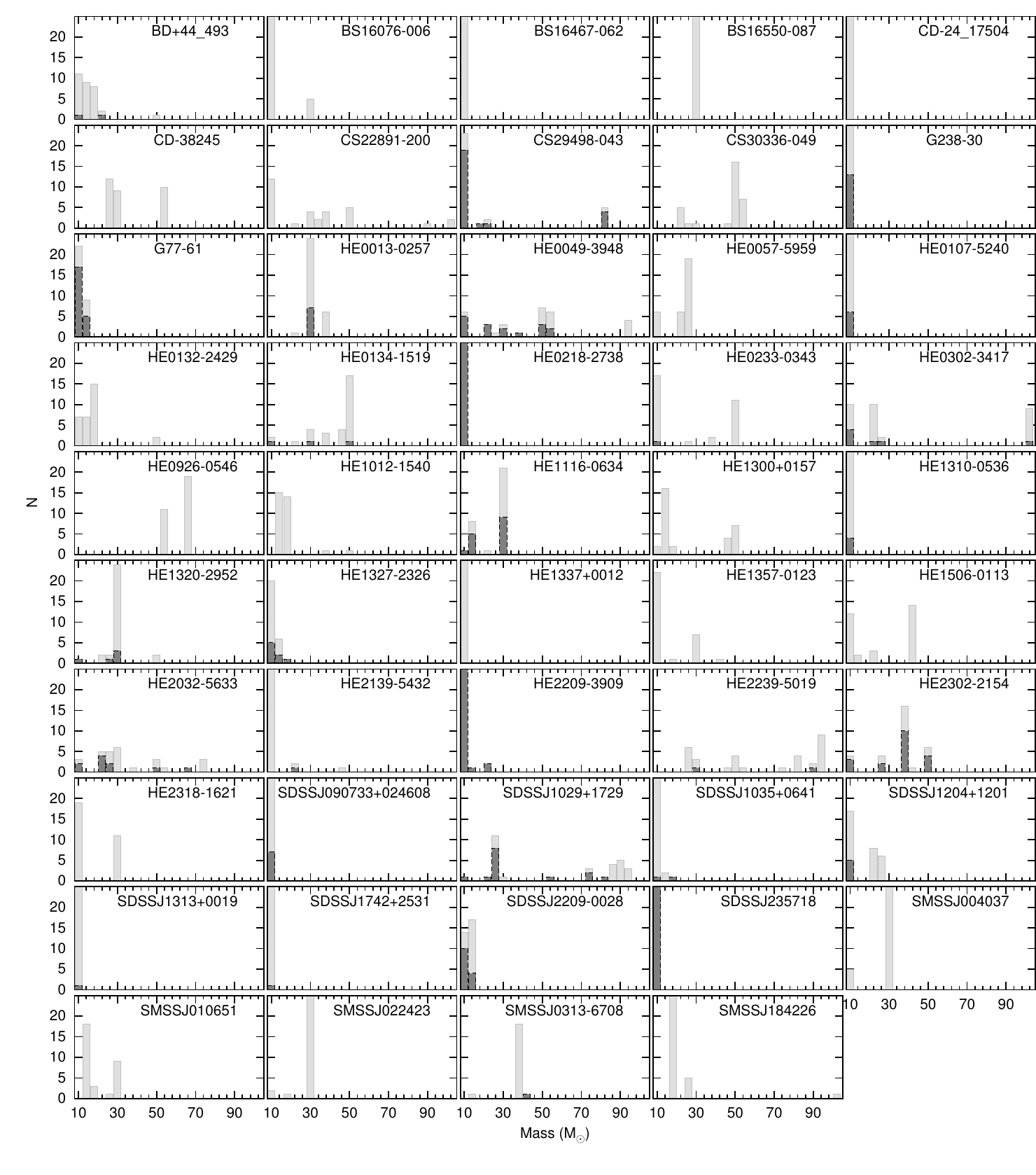}
   \caption{Histograms of derived masses from $30$ iterations of {\sc
       StarFit} using perturbed abundances. Light grey bars show fits
     with a residual of $<$25, while black lines show fits with a
     residual of $<$3.}
   \label{fig:histogram}
\end{figure*}

To quantify the uncertainty in derived progenitor masses, we perturbed
the measured abundances for each star using their quoted uncertainties
and queried {\sc StarFit} using the perturbed abundances.  We adopted
a conservative uncertainty of $0.2\,$dex for abundances without
reported uncertainties.  Measurements were assumed to be
normally-distributed, and upper limits were perturbed by drawing
values from a uniform distribution between ${\rm [X/H]} \sim
\mathcal{U}(-8, L)$, where $L$ is the published limit in [X/H] format.
We chose to draw from truncated uniform distributions for upper limits
as \citet{Placco15} notes low N abundances ([N/H]$ < -6$) are
consistently predicted by {\sc StarFit} when only upper limits on
nitrogen are available.  Draws in [X/H] were converted to
$\log_\epsilon({\rm X})$ before running {\sc StarFit}.  We repeated
this perturbation procedure thirty times for each star.  We find four
of the stars (HE0048-6408, HE0313-3640, HE2233-4724, and SMSSJ005953)
to have extremely large ($>25$) best-fit residuals for all
perturbations, and for this reason we discarded these stars from our
sample.  Numerical noise, uncertain physics \citep{Sukhbold_Woosley_2014}, 
and specifically the grid-like sampling of progenitor masses in supernova
models implies that there is a systematic uncertainty floor for a
given mass. Our tests
demonstrated that a 10 per cent uncertainty in progenitor masses was
sufficient to account for this sampling, which we take as
representative of the error on the fit.  
We account for the
distribution of recovered progenitor masses by means of a Monte Carlo
technique when fitting the IMF.  The distribution of inferred
progenitor masses for each star, after fitting our perturbed abundance
values are shown in Fig.~\ref{fig:histogram}
We stress that the error floor does not account for systematic differences 
between codes. While a detailed investigation of the differing yields from 
stellar evolution models is beyond the scope of this work, a comparison 
of solar metallicity models by \cite{Jon15} finds generally good agreement
among the GENEC, KEPLER and MESA codes up to the end of core He burning.
The differences at latter burning stages, and indeed in explosive nucleosynthesis
are harder to assess, and will depend on factors such as where the mass cut is placed
when exploding a model, and the degree of fallback after the initial SN collapse.
As such, our results should be regarded as model dependant.
We note however that for pair-instability SNe, the yields of the models computed 
by \cite{Koz14} compare favourably with those of \cite{Heg02}.

Sixteen of the stars in our sample also have SN progenitor mass
estimates from \cite{Placco15}.  \citeauthor{Placco15} use the same
{\sc StarFit} tool as this work, yet report slightly different
progenitor masses than what we find.  There are several resons for
these discrepancies. \citeauthor{Placco15} adjust the C abundances
based on the evolutionary state of the star, following empirical
corrections derived in \citet{Placco14}.  If no N measurement is
available, the authors also assume ${\rm [C/N]} = 0$.  In contrast, we
opt to fit the combined abundance of C, N, and O.  This option is
readily available in {\sc StarFit}, and implicitly accounts for CNO
cycling and the evolutionary state of the star.  Differences in
abundance choices also contributed to discrepancies in progenitor
masses.  We employed lower model limits for Cu and Sc (the default
recommended option), whereas \citeauthor{Placco15} quote model limits
for Cr and Sc.  Furthermore, for SMSS~J031300.36-670839.3, we opted for
the recommended abundances by \citet{Bessell15}, whereas
\citeauthor{Placco15} employed the 1D LTE abundances.  It also appears
that some abundances for elements with $Z < 30$ are not included in
the \citeauthor{Placco15} fits.  For example, the first star in Fig.~8
of \citeauthor{Placco15}, SDSS~J220924.70$-$002859.0, is missing both
the Na measurement and O upper limit from \citet{Spite13}.

The combination of these discrepancies and analysis choices explains
all differences in the inferred progenitor masses; by replicating the
\citeauthor{Placco15} \texttt{StarFit} options, adjusting C and
assuming [C/N]$ = 0$, and excluding abundances not present in their
Figure~8, we were able to reproduce all of their inferred progenitor
masses.  It is reassuring to see that even with the systematic
differences that arise from using different {\sc StarFit} settings,
the difference in inferred progenitor masses are reasonably
encapsulated by our total uncertainties.

We examined the distribution of derived masses from {\sc StarFit} as a
function of $T_\mathrm{eff}$, log~$g$, and [Fe/H].  There is no obvious
trend of progenitor mass with stellar parameters, which suggests that
our data are not introducing a systematic bias.  Encouragingly, we see
that of the three stars in our sample with precursor SN masses of
$>$70\msun, there is a spread in [Fe/H] between $-5.0$ and
$-3.6$~dex.

\begin{figure}
   \includegraphics[width=0.5\textwidth]{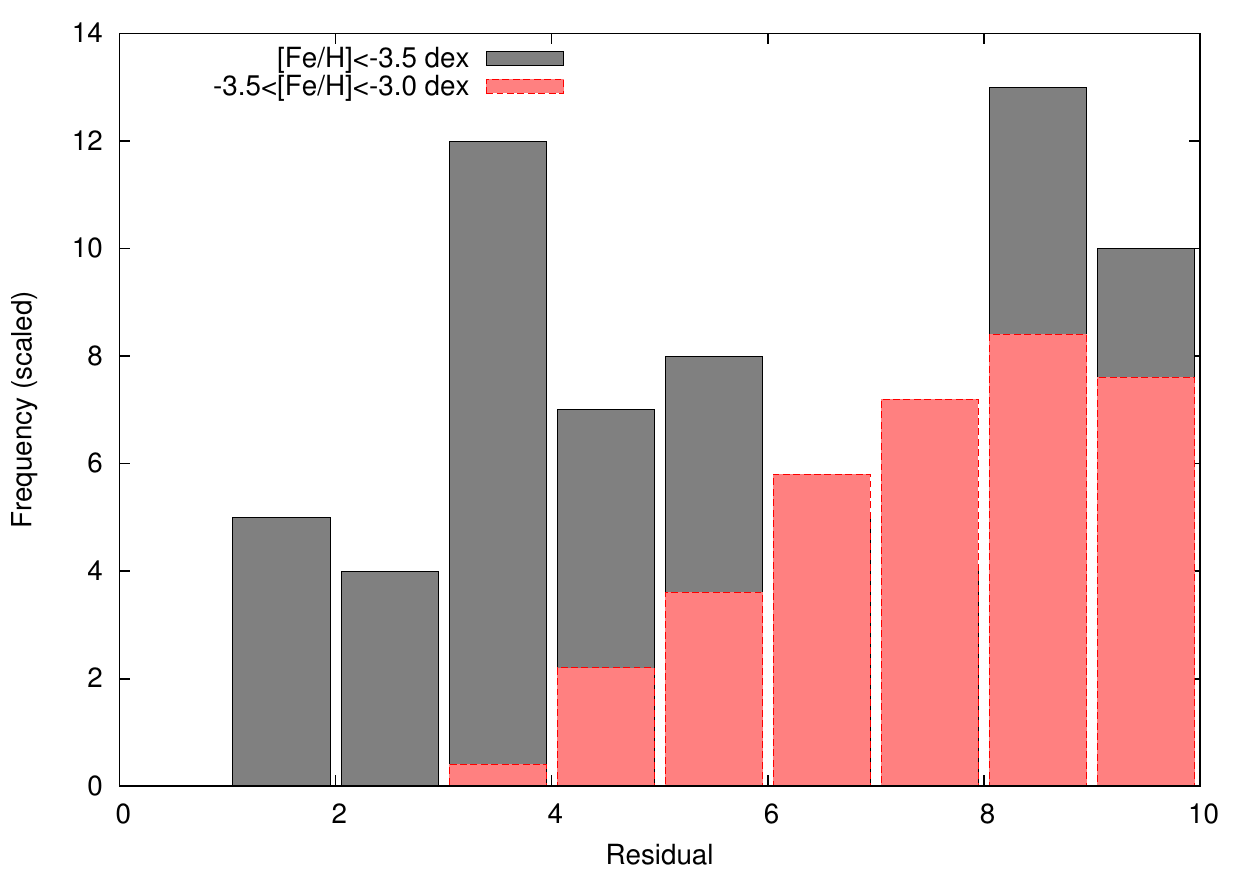}
   \caption{Distribution of residuals to the abundance fits, for stars above and below $-3.5$ dex. The lack of good fits to the higher metallicity stars is consistent with their being polluted by multiple SNe.}
   \label{fig:res}
\end{figure}

We also tested the effects of increasing our metallicity threshold from 
$-3.5$ to $-3.0$ dex. To do this, we took all stars with [Fe/H] between
$-3.5$ and $-3.0$ dex from \cite{Yong13}, and fitted their abundances as before.
We find that the residuals to the fit for the higher metallicity stars are larger. When
compared to the \cite{Heg02} models, no star above [Fe/H]$>$3.0 had a residual $<3$ (Fig. \ref{fig:res}).
This can be understood if these stars have been polluted by multiple SNe, and hence cannot
be fitted with single SN models. We note that even if we were to include these stars in our initial sample, their large
residuals would exclude them from the IMF fit described in the following.

\setcounter{table}{0}
\begin{table*}
 \begin{minipage}{\textwidth}
 \caption{Metal-poor star abundances used in this study.}
 \label{tab:emp-stars}
 \begin{tabular}{lcclllll}
 \hline
 \bf{Star} & \bf{$\alpha$ (J2000)} & \bf{$\delta$ (J2000)} & $\mathbf{T_{\rm eff}}$ & $\mathbf{\log{g}}$ & \bf{[Fe/H]} & $\mathbf{\xi}$ & \bf{Source} \\
 & \bf{[hms]} & \bf{[hms]} & \bf{[K]} & & & \bf{[km s$^{-1}$]} \\
\hline
BD+44~493                      & 02:26:49.7    & +44:57:46     & 5430 & 3.4   & $-$3.8  & 1.3   & \citet{Ito13} \\ %
BS~16076$-$006                 & 12:48:22.7    & +20:56:44     & 5199 & 3.0   & $-$3.81     & 1.4   & \citet{Bonifacio09} \\ %
BS~16467$-$062                 & 13:42:00.6    & +17:48:40     & 5388 & 3.04  & $-$3.7  & 1.7   & \citet{Lai08} \\ %
BS~16550$-$087                 & 14:10:26.7    & +18:01:23     & 4750 & 1.31  & $-$3.5  & 2.3   & \citet{Lai08} \\ %
CD$-$24~17504                  & 23:07:20.1    & $-$23:52:34     & 5821 & 3.5   & $-$3.66     & 1.22  & \citet{Ishigaki10} \\ %
CS~22891$-$200                 & 20:19:22.0    & $-$61:30:15     & 4500 & 0.45  & $-$3.92     & 2.6   & \citet{Hollek11} \\ 
CS~30336$-$049                 & 20:45:23.5    & $-$28:42:35     & 4725 & 1.19  & $-$4.1  & 2.1   & \citet{Yong13} \\ %
S~29498$-$043                 & 21:00:51.3    & $-$29:54:46     & 4400 & 0.6   & $-$3.75     & 2.3   & \citet{Aoki02} \\ %
G~77$-$61                      & 03:32:38.0    &   +01:58:00     & 4000 & 5.05  & $-$4.03     & 0.1   & \citet{Plez_Cohen05} \\ 
G~238$-$30                     & 13:17:40.2    &   +64:15:12     & 5299 & 3.39  & $-$3.72     & 1.19  & \citet{Ishigaki10} \\ %
HE~0013$-$0257                 & 00:16:04.2    & $-$02:41:06     & 4500 & 0.5   & $-$3.82     & 2.1   & \citet{Hollek11} \\ 
HE~0048$-$6408                 & 00:50:45.3    & $-$63:51:50     & 4378 & 0.15  & $-$3.75     & 2.85  & \citet{Placco14} \\ 
HE~0049$-$3948                 & 00:52:13.4    & $-$39:32:36     & 6466 & 3.78  & $-$3.68     & 0.8   & \citet{Yong13} \\ 
HE~0057$-$5959                 & 00:59:54.0    & $-$59:43:29     & 5257 & 2.65  & $-$4.08     & 1.5   & \citet{Yong13} \\ %
HE~0107$-$5240                 & 01:09:29.1    & $-$52:24:34     & 5100 & 2.2   & $-$5.3  & \nodata & \citet{Christlieb02} \\ %
HE~0132$-$2429                 & 01:34:58.8    & $-$24:24:18     & 5294 & \nodata & $-$3.55     &  \nodata & \citet{Cohen13} \\ 
HE~0134$-$1519                 & 01:37:05.4    & $-$15:04:24     & 5500 & 3.2   & $-$4.0  & 1.5   & \citet{Hansen14} \\ 
HE~0218$-$2738                 & 02:21:04.0    & $-$27:24:40     & 6550 & 4.3   & $-$3.54     & 0.84  & \citet{Carretta02} \\ %
HE~0233$-$0343                 & 02:36:29.7    & $-$03:30:06     & 6100 & 3.4   & $-$4.7  & 2.0   & \citet{Hansen14} \\ 
HE~0302$-$3417                 & 03:04:28.6    & $-$34:06:06     & 4400 & 0.2   & $-$3.7  & 2.0   & \citet{Hollek11} \\ 
HE~0313$-$3640                 & 03:15:01.8    & $-$36:29:54     & 6350 & 3.8 & $-$3.63 & 1.2 & \citet{Cohen13} \\ %
HE~0926$-$0546                 & 09:29:27.9    & $-$05:59:43     & 5159 & 2.50 & $-$3.73 & 1.8 & \citet{Cohen13} \\ %
HE~1012$-$1540                 & 10:14:53.5    & $-$15:55:54     & 5520 & 4.70 & $-$3.51 & 1.1 & \citet{Cohen13} \\ %
HE~1116$-$0634                 & 11:18:35.8    & $-$06:50:46     & 4400 & 0.1   & $-$3.73     & 2.4   & \citet{Hollek11} \\ 
HE~1300$+$0157                 & 13:02:56.3    & $+$01:41:51     & 5550 & 3.30 & $-$3.49 & 1.3  & \citet{Cohen13} \\ %
HE~1310$-$0536                 & 13:13:31.2    & $-$05:52:13     & 5000 & 1.9   & $-$4.2  & 2.2   & \citet{Hansen14} \\ 
HE~1320$-$2952                 & 13:22:54.9    & $-$30:08:05     & 5106 & 2.26  & $-$3.69 & 1.5    & \citet{Yong13} \\ 
HE~1327$-$2326                 & 13:30:06.0    & $-$23:41:51     & 6180 & 2.2   & $-$5.3  & \nodata & \citet{Frebel05} \\ 
HE~1337$+$0012                 & 13:40:02.4    & $-$00:02:18     & 6070 & 3.58  & $-$3.52     & 1.46  & \citet{Ishigaki10} \\ %
HE~1357$-$0123                 & 14:00:01.1    & $-$01:38:08     & 4600 & 1.05  & $-$3.8  & 2.1 & \citet{Cohen13} \\ %
HE~1506$-$0113                 & 15:09:14.3    & $-$01:24:56     & 5016 & 2.01  & $-$3.54 & 1.6    & \citet{Yong13} \\ 
HE~2032$-$5633                 & 20:36:24.9    & $-$56:23:05     & 6457 & 3.78  & $-$3.63 & 1.8     & \citet{Yong13} \\ 
HE~2139$-$5432                 & 21:42:42.4    & $-$54:18:42     & 5416 & 3.04  & $-$4.02     & 0.8   & \citet{Yong13} \\ %
HE~2209$-$3909                 & 22:12:00.7    & $-$38:55:02     & 6305 & 4.3   & $-$3.66     & 1.2 & \citet{Cohen13} \\ %
HE~2233$-$4724                 & 22:35:59.2    & $-$47:08:36     & 4360 & 0.4   & $-$3.65     & 3.0   & \citet{Placco14} \\ 
HE~2239$-$5019                 & 22:42:26.9    & $-$50:04:01     & 6100 & 3.5   & $-$4.2  & 1.8   & \citet{Hansen14} \\ 
HE~2302$-$2154                 & 23:05:25.2    & $-$21:38:07     & 4675 & 0.9   & $-$3.9  & 2.0   & \citet{Hollek11} \\ 
HE~2318$-$1621                 & 23:21:21.5    & $-$16:05:06     & 4846 & 1.4   & $-$3.67     & 1.75  & \citet{Placco14} \\ 
SDSS~J090733.30$+$024608.0     & 09:07:33.3    & +02:46:08     & 5934 & 4.0   & $-$3.52     & 1.8   & \citet{Caffau11a} \\ 
SDSS~J102915.15$+$172928.0     & 10:29:15.2    & +17:29:28     & 5811 & 4.0   & $-$4.99     & 1.5   & \citet{Caffau11b} \\ 
SDSS~J103556.10$+$064143.0     & 10:35:56.1    & +06:41:43     & 6262 & 4.0   & $<-$5.07    & 1.5   & \citet{Bonifacio15} \\ 
SDSS~J1204$+$1201              & $\sim{}$12:04 & $\sim$+12:01  & 5467 & 3.2   & $-$4.34     & 1.5   & \citet{Placco15} \\ %
SDSS~J131326.90$+$001941.0     & 13:13:26.9    & $-$00:19:41     & 5200 & 2.6   & $-$5.0  & 1.8   & \citet{Frebel15} \\ 
SDSS~J174259.70$+$253135.0     & 17:42:59.7    & +25:31:35     & 6345 & 4.0   & $-$4.8  & 1.5   & \citet{Bonifacio15} \\ %
SDSS~J220924.70$-$002859.0     & 22:09:24.7    & $-$00:28:59     & 6440 & 4.0   & $-$4.0  & 1.3   & \citet{Spite13} \\ %
SDSS~J235718.91$-$005247.8     & 23:57:18.9    & $-$00:52:47     & 5000 & 4.8   & $-$3.52     & 0.0   & \citet{Aoki10} \\ %
SMSS~J004037.56$-$515025.2     & 00:40:37.5    & $-$51:50:25     & 4468 & 1.05  & $-$3.67     & 2.45  & \citet{Jacobson15} \\ %
SMSS~J005953.98$-$594329.9     & 00:59:53.9    & $-$59:43:29     & 5413 & 3.41  & $-$3.78     & 1.4   & \citet{Jacobson15} \\ %
SMSS~J010651.91$-$524410.5     & 01:06:51.9    & $-$52:44:10     & 4486 & 1.15  & $-$3.63     & 2.6   & \citet{Jacobson15} \\ %
SMSS~J022423.27$-$573705.1     & 02:24:23.2    & $-$57:37:05     & 4846 & 2.33  & $-$3.74     & 2.05  & \citet{Jacobson15} \\ %
SMSS~J031300.36$-$670839.3     & 03:13:00.3    & $-$67:08:39     & 5125 & 2.3   & $<-$7.52    & 2.0   & \citet{Bessell15} \\ 
SMSS~J184226.25$-$272602.7     & 18:42:26.2    & $-$27:26:02     & 4450 & 1.25  & $-$3.73     & 2.85  & \citet{Jacobson15} \\ %
\hline
 \end{tabular}
 \end{minipage}
\end{table*}

\begin{figure}
   \includegraphics[width=0.38\textwidth,angle=90]{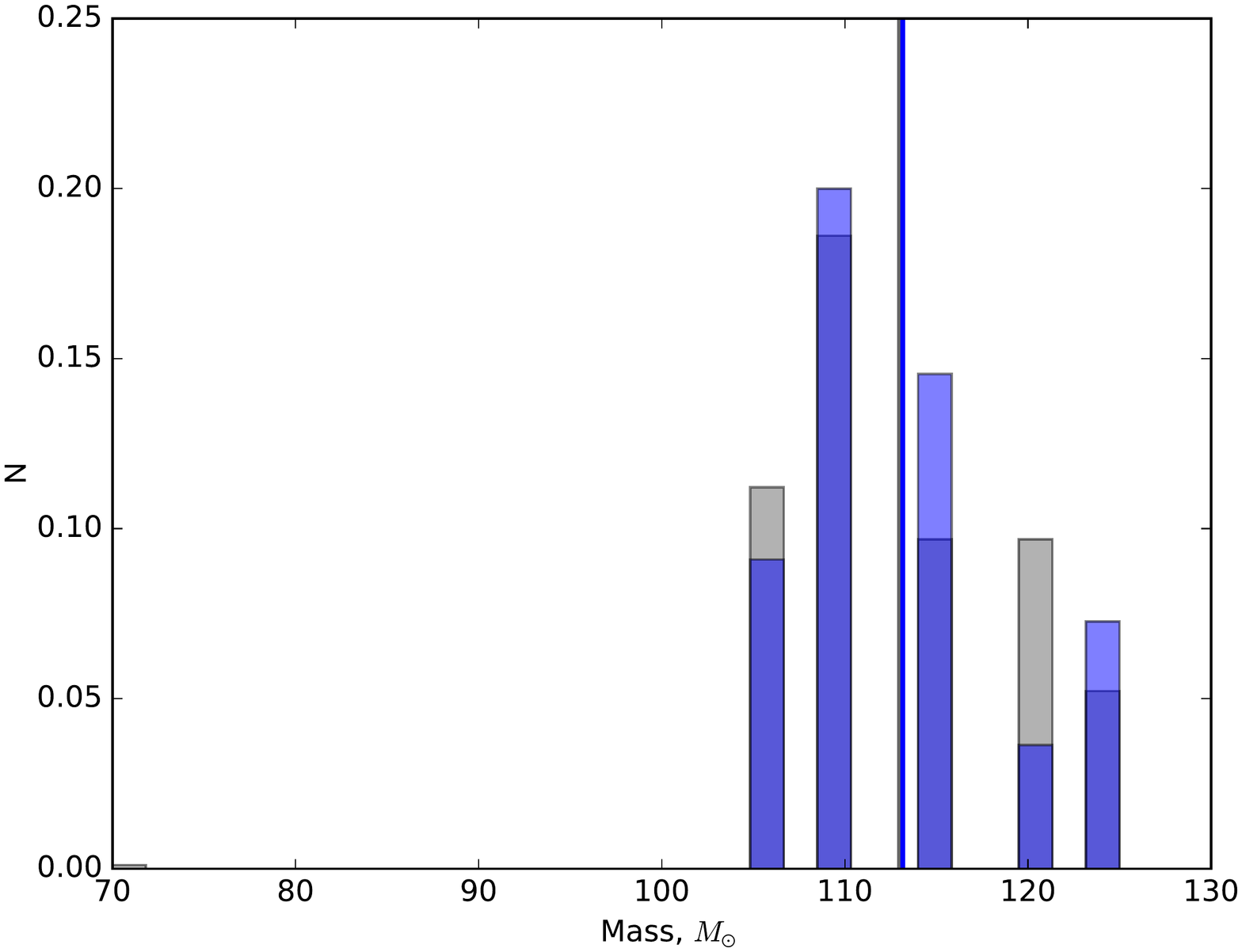}
   \caption{The normalised distribution of masses from 500 sets of perturbed abundances for a single star (grey) compared the distribution from 30 perturbed abundances (blue) and normalized histograms of mass for one star (grey = 500 trials, blue = 30). The thin line marks the mean of each set of trials. Note that no cut has been applied to the fits based on their residuals.}
   \label{fig:trials}
\end{figure}

\begin{figure*}

   \includegraphics[width=0.75\textwidth]{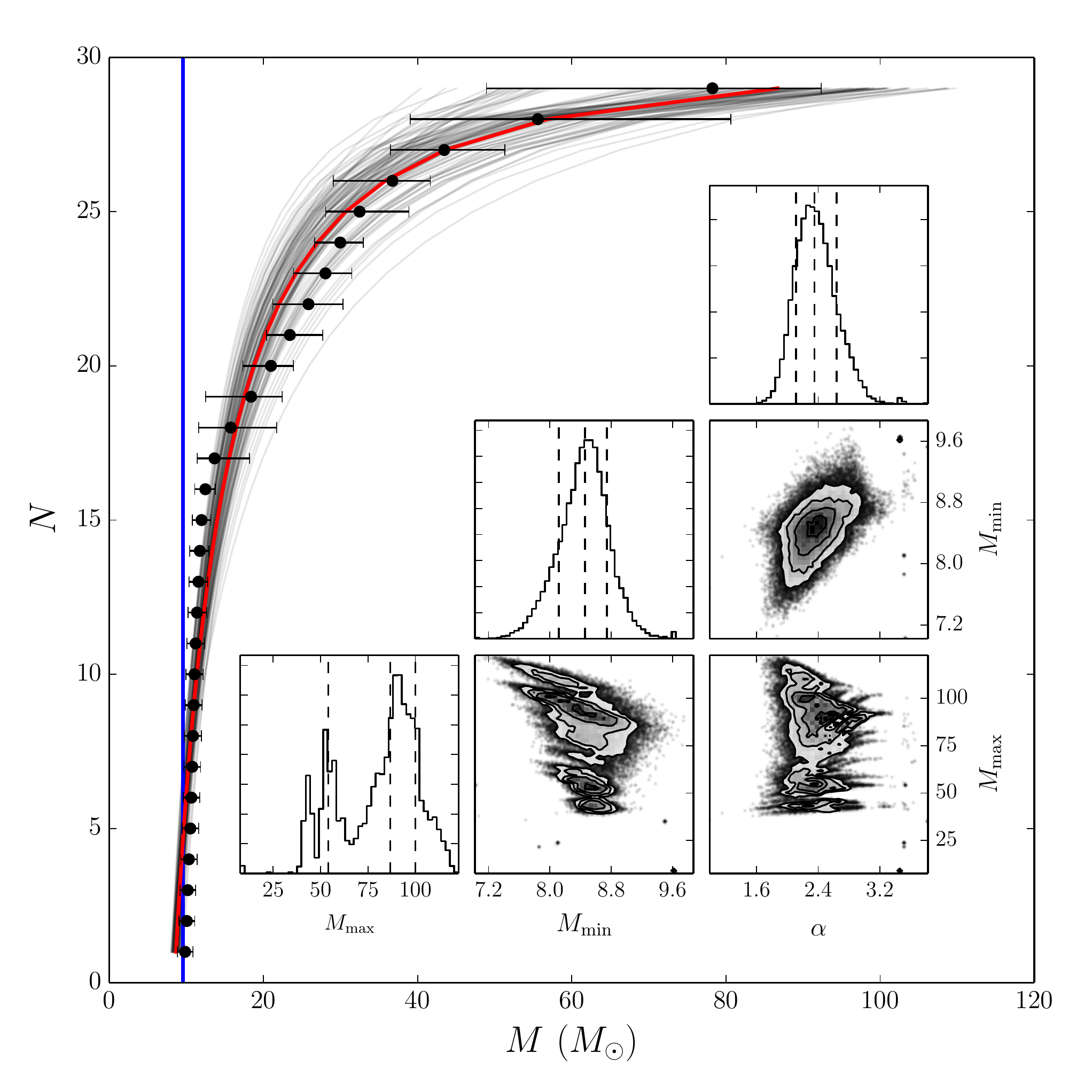}
   \caption{Main panel: The cumulative distribution of Population III
     supernova progenitor masses to which the IMF was fit.  Black
     points show the mean mass of the $n$th star across all
     Monte-Carlo iterations.  Error bars correspond to the 16th and
     84th percentile ranges (i.e., $\simeq1\sigma$ of the mass of the
     $n$th star, added in quadrature with the 10 percent error
     floor.)  The light grey lines show 100 IMFs as fitted during
     Monte-Carlo iterations, while the red line show the final best
     fit IMF.  The blue vertical line shows the lower mass limit of 9.6 \msun\ for the
     model grid used with {\sc starfit}.  Inset panel: Best fit values
     for \alf, \mmin\ and \mmax\ from 50,000 Monte-Carlo
     iterations. The distribution of values for each parameter is
     shown with a histogram, and the best fit value and 16th and 84th
     percentile limits are marked with dashed lines. \mmin\ and \mmax\ are in 
     units of \msun.}
   \label{fig:fit}
\end{figure*}

The Initial Mass Function can be defined as ${\xi(M)=AM^{-\alpha}}$
where $\xi(M)$ is the probability of finding a star of mass $M$, the
exponent $\alpha$ is the ``slope'' of the IMF, and $A$ is a
normalisation constant \citep{Sal55}.  To determine the number of
stars between masses $M_1$ and $M_2$, we simply integrate with respect to M, over the range $M_1$ to $M_2$.
When fitting the observed distribution of Population III supernova
progenitor masses to the IMF, we have three free parameters: $\alpha$,
and the minimum and maximum mass for a SN progenitor, $M_{\rm min}$
and $M_{\rm max}$ respectively.  As can be seen from
Fig.~\ref{fig:histogram}, the masses derived for each progenitor when
perturbing abundances are not normally distributed.  We hence employed
a Monte Carlo technique, where we fit the IMF to our observed data
50,000 times.  For each iteration, we randomly draw a mass for each star
from the 30 perturbed measurements, discarding any masses where {\sc
  StarFit} returned a residual of greater than 3 when fitting
abundances.
For a single star, we also tested the distribution of
masses obtained from 500 perturbed measurements, and encouragingly found this to be compatible with 
that obtained from 30 (Fig. \ref{fig:trials}).
The parameters of the IMF from each Monte Carlo iteration
were simply taken as those values which minimised the $\chi^2$ value
of the fit, as shown in Fig.~\ref{fig:fit}.  The constraint that the
residual of the {\sc starfit} model must be $<$3 reduces the size of
the sample from 49 stars to 29.

Given the progenitor masses in shown in Fig.~\ref{fig:histogram}, the
best fit values (and 16th/84th percentile uncertainties) we infer for
the IMF are ${\alpha=2.35^{+0.29}_{-0.24}}$, a formal minimum SN
progenitor mass of ${M_\mathrm{min}=8.5_{-0.4}^{+0.2}}$\msun\ and a
maximum SN progenitor mass of ${M_\mathrm{max}=87_{-33}^{+13}}$\msun.
Since the minimum mass for Population III supernovae in the data base is
9.6\msun, however, the conclusion is that the lower IMF endpoint must
lie below the minimum mass for Population III supernovae.  From the present
data, no conclusion can be drawn, however, as to how far below that may be.

\section{Discussion and conclusions}
\label{sect:discussion}

Overall, the fit to the inferred Population III supernova progenitor
masses found in Sect.~\ref{sect:fitting} appears quite good.  In
detail, there are three mass ranges where the quality of the fit
appears to vary.  Above 30 \msun\ the inferred masses match the fitted
IMF extremely well.  We find that the distribution of possible maximum
masses for a Population III supernova progenitor to be be roughly
bimodal, with a main peak at $\sim$90 \msun, and a smaller peak at
$\sim$50 \msun.  There is a small tail of Monte Carlo trials which
accommodate an upper mass limit to our sample of $\lesssim$120 \msun, but none above
this.  We also note that the upper mass limit is relatively insensitive
to $\alpha$.

Between 15 \msun\ and 30 \msun, the inferred progenitor mass
distribution appears to lie slightly below the best fit IMF.  The
models to which we fit the abundances have masses in steps of at most
0.5 \msun\ over this mass range, so grid effects are unlikely to be
the cause of this.  An alternative possibility is that an appreciable
fraction of stars with zero-age main sequence masses in the range
15--30 \msun\ massive stars do not explode as SN but make black holes
instead \citep{Suk14}, and hence do not eject their synthesised
material to enrich subsequent generations of star formation.

Below 15 \msun\, the inferred masses appear to follow a somewhat
steeper IMF than our best fit, suggesting that the exponent $\alpha$
may be slightly higher.  The IMF is pulled towards shallower slopes
(i.e., smaller values of $\alpha$) by the stars in the 15 -- 30
\msun\ range.  If these were excluded, then the fit could better
accommodate the lower mass stars, although we are wary of arbitrarily removing
stars to improve the fit.  The formally deduced minimum mass
of $8.5^{+0.2}_{-0.4}$\msun\ from statistical sampling lies well below
the minimum mass for Population III supernovae as reflected by stars in the
model data base. We can infer that there is no evidence
that the minimum mass for a Population III SN was significantly higher
than eight solar masses. However, as the lower mass limit 
is close to the edge of the model grid 
at 10\msun\, we stress that we cannot determine from this data 
whether the IMF for all stars (and not just those that explode) extends to much lower masses.

\begin{figure}
   \includegraphics[width=0.5\textwidth]{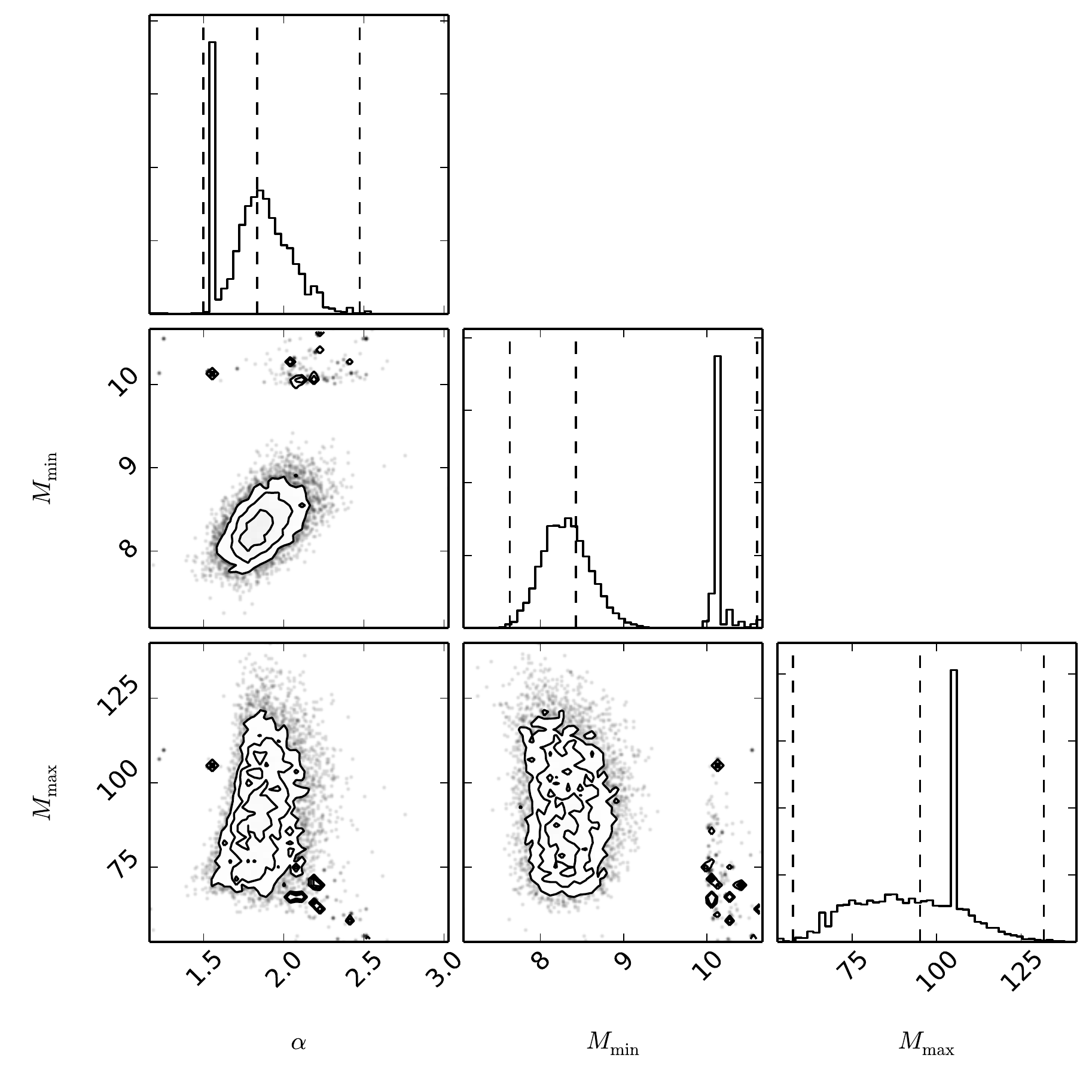}
   \caption{The best fit values of \alf, \mmin\ and \mmax\ from Monte-Carlo
     trials, fitting the 46 stars which had derived SN progenitor masses with 
     residuals from {\sc starfit} $<$10. The distribution of values for each parameter is
     shown with a histogram, and the best fit value and 16th and 84th
     percentile limits are marked with dashed lines. \mmin\ and \mmax\ are in 
     units of \msun.}
   \label{fig:fit_less10}
\end{figure}

As a test, we also relaxed our constraint that the residual of the
{\sc starfit} models must be $<$3, and allowed for models with a
residual of up to 10. 46 stars had acceptably fitted masses, and these
gave a lower limit of $8.3_{-0.7}^{+0.8}$ \msun, and an upper mass limit of
between 70 and 110 \msun\ (with no clear peak in the distribution),  as shown in Fig. \ref{fig:fit_less10}.  Again, a contribution from stars above 125
\msun\ can be excluded at high significance, although interestingly
the fit prefers a flatter IMF slope of $\alpha = 1.8_{-0.3}^{+0.7}$ (which is however still formally consistent with the result from stars with residual $<3$).

We find the best fit value of the exponent in the IMF to be 2.35,
which is identical to the canonical value of \cite{Sal55}, and perhaps
suggests that star formation in the metal free Universe is not so
dissimilar to that in the present day.  We see no evidence for stars
above $\sim$120\msun\ exploding  in our sample and no signs of the hypothesised
pair-instability SNe which may have enriched the Universe at the
earliest times.  However, we caution that our sample is small, and 
with 29 stars drawn from a Salpeter IMF between 8\msun\ and 300\msun\ 
we would have only a one in three chance of seeing a star above 150\msun.
Hence we cannot set strong limits on the existence of very massive stars which 
exploded as pair instability SNe with the current dataset. However, increasing the size 
of our sample of EMP stars will allow for stronger constraints to be placed on population
III SN progenitors in the future.

 For comparison, in the present, approximately solar metallicity local Universe, 
direct detections of core-collapse supernova progenitors in a volume limited sample 
indicate that most of these are stars have masses $\lesssim$ 17\msun\ \citep{Sma15}.
Similar conclusions can be drawn from modelling of nucleosynthetic yields of individual SNe \citep{Jer14}.
This apparent dearth of relatively high mass SN progenitors in nearby galaxies
stands in contrast to what we infer in this paper for Population III SN progenitors.

We note that the fact that perturbing abundances within their
quoted uncertainties affects the inferred progenitor mass to such a
large extent (as shown in Fig.~\ref{fig:histogram}) demonstrates the
necessity of propagation or sampling techniques to account for the
full distribution of progenitor masses.

Finally, we emphasize that our abundance probe is only a reflection of
the fraction of stars that do explode as SNe; we have no probe of
stars that don't explode because they make black holes or are too low mass.  Variation in explosion energy, the amount
of matter that is enriched by
supernovae of a given mass, and the potentially environment-dependent
fraction of such gas forming low-mass stars, may also affect the
conclusions drawn above and will require further extended studies of
first star formation and death.

\section*{Acknowledgments}

We thank the anonymous referee for their helpful and constructive comments.
This work was supported by the European Union FP7 programme through
ERC grant number 320360. MF is supported by a Royal Society - Science Foundation
Ireland University Research Fellowship.
AH was supported
by Australian Research Council through a Future Fellowship (FT120100363). This research has made use of the SIMBAD
database, operated at CDS, Strasbourg, France, the SAGA database
\citep{Suda08}, NASA's Astrophysics Data System, the excellent {\sc
  triangle} package \citep{triangle}, and Astropy \citep{astropy}.

\def \apjl {ApJL}
\def \aap {A\&A}
\def \nat {Nature}
\def \aj {AJ}
\def \apj {ApJ}
\def \pasj {PASJ}
\def \mnras {MNRAS}
\def \araa {ARAA}
\def \apjs {ApJS}
\def \aaps {A\&AS}
\def \astro {astro-ph}

\label{lastpage}

\end{document}